\documentstyle[12pt]{article}
\begin{document}
\begin{center}
{\bf {\Large Phase properties of hypergeometric states and negative hypergeometric states\\[0pt]
}} 
\baselineskip=25pt \vspace{2cm} Xiao-Guang Wang\footnote{%
email:xyw@aphy.iphy.ac.cn}\\[0pt]
{Laboratory of Optical Physics, Institute of Physics,\\[0pt]
Chinese Academy of Sciences,Beijing,100080,P.R.China\\[0pt]
} 

\end{center}

\vspace{1cm} \baselineskip=20pt {\bf Abstract.} 

We show that the three quantum states (P$\acute{o}$lya states, the
generalized non-classical states
related to Hahn polynomials and negative hypergeometric states)
introduced recently as intermediates states which interpolate between the
binomial states and negative binomial states are essentially identical.
By using the Hermitial-phase-operator formalism, the phase properties
of the hypergeometric states and negative hypergeometric
states are studied in detail. We find that the number of peaks of phase probability distribution is one for the hypergeometric states and $M$ for
the negative hypergeometric states.

\vspace{0.5cm} {\bf PACS numbers:}42.50.Dv,03.65.Db,32.80.Pj,42.50.Vk

\newpage

\section{Introduction}
In 1985, Stoler et al. introduced the binomial states (BSs) of
a single mode of electromagnetic field[1]. The BS is defined as
\begin{equation}
|\eta,M\rangle=\sum_{n=0}^{M}\left[{M\choose n}\eta^n(1-\eta
)^{M-n}\right]^{1/2}|n\rangle.
\end{equation}
Here $|n\rangle$ is the usual number state, $\eta$ is a real parameter satisfying $0\le\eta\le 1$ and
$M$ is a non-negative integer. Some of their properties[1-6],
methods of generation[1-3] as well as their interacting with atoms
[7] have already been discussed in the literature. The BSs are intermediate
number-coherent states in the sense that they reduce to the number
and coherent states in two different limits. Complementary to
the BSs, the negative binomial states(NBSs) are also introduced
and studied[8-13], they interpolate between the Susskind-Glogower
phase states[14] and coherent states.

Recently three quantum states, P$\acute{o}$lya states(PSs)[15],the generalized non-classical states related to Hahn polynomials[16], and negative hypergeometric states(NHGSs)[17]
are introduced as different intermediate BS-NBS states. They seem different,
but are essentially identical. We will prove this in the next
section.

The hypergeometric states(HGSs) which are complementary to
the NHGSs are introduced[18] and defined as
\begin{equation}
|L,M,\eta\rangle=\sum_{n=0}^M H_n^M(\eta,L)|n\rangle=\sum_{n=0}^M\left[{L\eta\choose n}{L\bar{\eta}\choose M-n}\right]^{1/2}{L\choose M}^{-1/2}|n\rangle,
\end{equation} 
where  $\bar{\eta}=1-\eta$  , $L$ is a real number
satisfying $L\ge \max\{M\eta^{-1},M\bar{\eta}^{-1}\}$, and 
\begin{equation}
{x\choose n}=\frac{x (x-1)...(x-n+1)}{n!}, {x\choose 0}\equiv 1.
\end{equation}
Note that in Eq.(3)
the real number $x$ is not necessarily an integer.
The HGS can reduce to the BS in certain limit and the BS to
the number and coherent state.

In Section 3 of this article, we will study
the phase properties of the HGS and NHGS based on
the Hermitian-phase-operator formalism.
A conclusion is given in Section 4.
\section{Equivalence of the three intermediate BS-NBS states}

The P$\acute{o}$lya state is defined as[16]
\begin{equation}
|M,\gamma,\eta\rangle=\sum_{n=0}^{M}P_n^M(\gamma,\eta)|n\rangle,
\end{equation}
where 
\begin{eqnarray}
P_n^M(\gamma,\eta)&=&{M\choose n}^{1/2} \{\prod_{k=1}^n[\eta+(k-1)\gamma]\}^{1/2}\{\prod_{k=1}^{M-n}[\bar{\eta}+(k-1)
\gamma]\}^{1/2}\nonumber\\
&&\{\prod_{k=1}^M[1+(k-1)\gamma]\}^{-1/2}
\end{eqnarray}
and $\gamma>0$ is a real constant.
The generalized non-classical state related to Hahn polynomials is defined as[17]
\begin{equation}
|\alpha,\beta,M\rangle=\sum_{n=0}^{M}{M\choose n}^{1/2}[(\alpha+1)_n(\beta+1)_{M-n}(\alpha+\beta+2)_M^{-1}]^{1/2}
|n\rangle,
\end{equation}
where $\alpha, \beta>-1$ are real numbers and 
\begin{equation}
(x)_n=x(x+1)...(x+n-1), (x)_0\equiv 1.
\end{equation}

Both the PS and generalized non-classical state can degenerate to the BS and 
NBS. This promotes us to investigate the relation between them. In fact, the 
coefficients of the PS can be written as
\begin{equation}
P_{n}^{M}(\gamma,\eta)={M\choose n}^{1/2}
\left[(\eta/\gamma)_n(\bar{\eta}/\gamma)_{M-n}
(1/\gamma)^{-1}_{M}\right]^{1/2}.
\end{equation}
Setting
\begin{equation}
\eta/\gamma=\alpha+1, \bar{\eta}/\gamma=\beta+1,
\end{equation}
we find the coefficients of the PS are identical
to those of the generalized non-classical state
related to Hahn polynomials. Although the two states show
different forms, they are really equivalent.

The negative hypergeometric state introduced by Fan et al. is defined as
[18]
\begin{equation}
|M,\beta,s\rangle=\sum_{n=0}^{M}\Theta_n^M(\beta,s)|n\rangle,
\end{equation}
where
\begin{equation}
\Theta_n^M(\beta,s)=\left[{n+s\choose n}{M/(1-\beta)-n-s-1\choose M-n} {M/(1-\beta)\choose M}^{-1}\right]^{1/2},
\end{equation}
$\beta$ is real number and $s$ a non-negative integer satisfying 
$s<M\beta/(1-\beta)<M/(1-\beta)$.
The NHGS is also claimed to be a intermediate BS-NBS state. We guess that it is equivalent
to the PS and the generalized non-classical state.

Using the following identities
\begin{eqnarray}
{n+s\choose n}&=&\frac{(1+s)_n}{n!},\nonumber\\
{M/(1-\beta)-n-s-1\choose M-n}&=&\frac{(M\beta/(1-\beta)-s))_{M-n}}{(M-n)!},\nonumber\\
{M/(1-\beta)\choose M}&=&\frac{(M\beta/(1-\beta)+1)_M}{M!},
\end{eqnarray}
we write the coefficients of the NHGS as
\begin{equation}
\Theta_n^{M}(\beta,s)={M\choose n}^{1/2}\left[(s+1)_n(M\beta/(1-\beta)-s)_{M-n}(M\beta/(1-\beta)+1)^{-1}
\right]^{1/2}.
\end{equation}
Comparing Eq.(8) and Eq.(13) and setting 
\begin{equation}
\eta/\gamma=s+1, \bar{\eta}/\gamma=M\beta/(1-\beta)-s,
\end{equation}
we find that the NHGS and PS are equivalent. Thus we conclude
that the three intermediate BS-NBS states, the PS, the generalized non-classical
state related to Hahn polynomials and the NHGS are equivalent.
There is only one intermediate BS-NBS state other than three
up to now.
In the following, we adopt the form of  PS to discuss the phase properties.

\section{Phase properties of HGS and NHGS}
\subsection{Hermitian-phase-operator formalism}
Pegg and Barnett[19-22]defined the Hermitian phase operator on a finite-dimensional state space, which make it possible to study the phase properties of quantum states of the single mode of the electromagnetic field. 
The phase states $|\theta_m\rangle$ can be defined as
\begin{equation}
|\theta_m\rangle=\frac{1}{(s+1)^{1/2}}\sum_{n=0}^{s}\exp(in\theta_m)|n\rangle
\end{equation}
The phase states in Eq.(15) form an orthonormal set provided that we have
\begin{equation}
|\theta_m\rangle=\theta_0 + 2\pi m/(s+1), m=0,1,...,s,
\end{equation}
where $\theta_0$ is an arbitrary reference phase. In order to be consistent with quantum mechanical formalism, we must take the limit $s\rightarrow \infty$ after all the calculations. A Hermitian phase operator can be constructed using the phase states themselves:
\begin{equation}
\Phi_\theta=\sum_{m=0}^{s}\theta_m|\theta_m\rangle\langle\theta_m|.
\end{equation}
It is not difficult to obtain expressions for the phase variance and phase distribution when the quantum state is a partial phase state defined by
\begin{equation}
|b\rangle=\sum_{n=0}^{s} b_n \exp(in\mu)|n\rangle,
\end{equation}
where $b_n$ is real and positive.  From Eq.(17) and (18), the mean phase is obtained as
\begin{equation}
\langle b|\Phi_\theta|b\rangle=\mu.
\end{equation}
The phase variance is given by
\begin{equation}
\langle\Delta\Phi_\theta^2\rangle=\frac{\pi^2}{3}+4\sum_{n>n'}b_nb_{n'}(-1)^{n-n'}(n-n')^{-2}
\end{equation}
and the phase probability distribution can be written as
\begin{equation}
P(\theta)=\frac{1}{2\pi}\left\{1+2\sum_{n>n'}b_nb_{n'}\cos[(n-n')\theta)]
\right\},
\end{equation}
after taking the limit $s\rightarrow\infty$.
\subsection{Phase properties}
Before investigating the phase properties, we would like to remark that in
the case of $M=1$, the HGS is $L$ independent and the NHGS is $\gamma$ 
independent. This is  because the HGS and NHGS are equal to the 
binomial state for $M=1$.
The HGS and NHGS are in fact partial phase states
according to the definitions of the partial phase states.
So we immediately have, from Eq.(19), that the mean phase in a hypergeometric state or negative hypergeometric state is always zero.
Using Eqs.(2),(4),(20) and (21), we can investigate
phase properties of the HGS and NHGS.

In Fig.1 we have given the variances of the phase operator of a HGS for 
different values of $M$. The variance for
the case of $M=1$ is independent of $L$. It is understandable from the above remark. It can be seen that the variance decreases as $M$ increases. Fig.2 is a plot of the phase probability distributions of a HGS as a function of $\theta$ for different values of $L$. It is known that the HGS can degenerate to the BS in the limit
$L\rightarrow\infty$[18]. For $L=1200$,
the phase probability of the HGS
$|L,\eta,M\rangle$ is close to that of the BS $|\eta,M\rangle$. 
The maxima of the phase probability
distributions of the HGS is always lower than
that of the corresponding BS.
The number of the maximum of a HGS is only one 
and the value of the maximum increases as $L$ increases.

Fig.3 is plot of the variances of the phase operator of a NHGS as a function
of $\eta$ for different values of $\gamma$. The variance first decreases as $\eta$ increases, having a minimum at $\eta=0.5$, and there is an increase
up to $\eta=1$. The larger $\gamma$, the larger the minimum value.
The phase probability distributions of a NHGS are shown in Fig.4. The distributions are symmetric about the point $\theta=0$.
The number of peaks of the distribution is equal to $M$.
 
\section{Conclusion}
We have studied the HGS and NHGS and their phase properties.
A fact is clarified that the three intermediate BS-NBS states introduced
recently by three different authors are essentially identical.
Up to now there is only one intermediate BS-NBS state. We can call the 
intermediate state as negative hypergeometric state
in order to be complementary to the HGS just as
the BS to the NBS.
The variances and phase probability distributions are studied based on 
the Hermitian-phase-operator formalism. It is interesting that the number of 
peaks of the phase distribution is one for the HGS and $M$ for the NHGS.

\vspace{1cm} {\bf Acknowledgment}: The author thanks for the
discussions with Prof. H.C.Fu and the help of Prof. C.P.Sun, S.H.Pan
and G.Z.Yang. The work is
partially supported by the National Science Foundation of China
with grant
number:19875008.

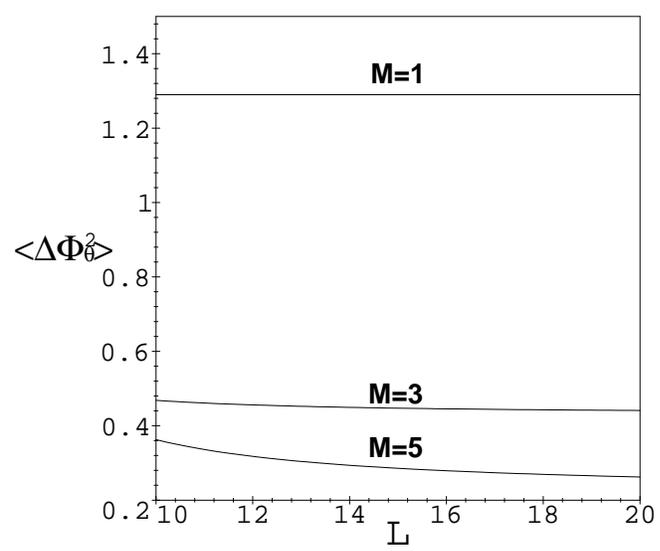
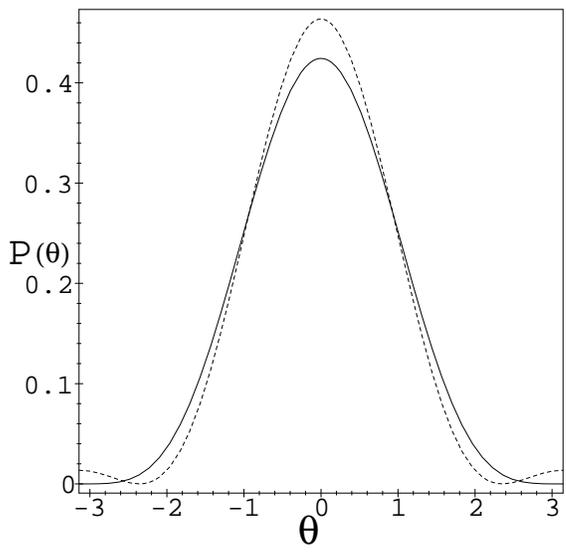
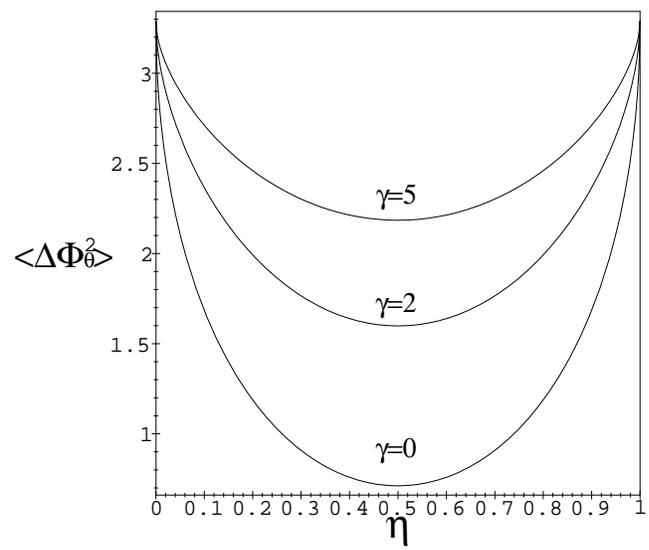
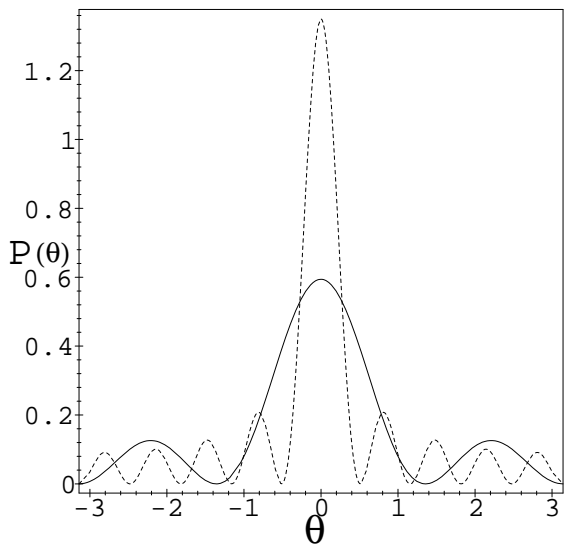
\end{document}